\documentclass[dvipsnames,format=sigconf,anonymous=false,review=false]{acmart}

\AtBeginDocument{%
  \providecommand\BibTeX{{%
    \normalfont B\kern-0.5em{\scshape i\kern-0.25em b}\kern-0.8em\TeX}}}

\setcopyright{acmcopyright}
\copyrightyear{2023}
\acmYear{2023}
\acmDOI{10.1145/1122445.1122456}

\acmConference[GECCO '23]{Genetic and Evolutionary Computation Conference Companion (GECCO’23 Companion)}{July 15--19, 2023}{Lisbon, Portugal}
\acmBooktitle{Genetic and Evolutionary Computation Conference Companion (GECCO’23 Companion),
  July 15--19, 2023, Lisbon, Portugal}
\acmPrice{15.00}
\acmISBN{978-1-4503-XXXX-X/18/06}

\usepackage{blindtext}
\usepackage{braket}
\usepackage{nicefrac}
\usepackage{mathtools}

\usepackage{tikz}
\usetikzlibrary{quantikz}
\tikzset{slice/.append style={line width=1.5pt}}

\usepackage{amsmath}
\usepackage{outlines}
\usepackage{mathtools}
\usepackage{graphicx}
\usepackage{caption,mwe}
\usepackage{subcaption}

\begin{document}

\title{Evidence that PUBO outperforms QUBO when solving continuous optimization problems with the QAOA}

\author{Jonas Stein}
\email{jonas.stein@ifi.lmu.de}
\orcid{0000-0001-5727-9151}
\affiliation{%
  \institution{LMU Munich}
  \streetaddress{Oettingenstr. 67}
  \city{Munich}
  \state{Bavaria}
  \country{Germany}
  \postcode{80538}}

\author{Farbod Chamanian}
\orcid{0009-0002-3027-8241}
\affiliation{%
  \institution{LMU Munich}
  \streetaddress{Oettingenstr. 67}
  \city{Munich}
  \state{Bavaria}
  \country{Germany}
  \postcode{80538}}
\email{farbod.ch.96@gmail.com}

\author{Maximilian Zorn}
\affiliation{%
  \institution{LMU Munich}
  \streetaddress{Oettingenstr. 67}
  \city{Munich}
  \state{Bavaria}
  \country{Germany}
  \postcode{80538}}
\email{maximilian.zorn@ifi.lmu.de}

\author{Jonas Nüßlein}
\affiliation{%
  \institution{LMU Munich}
  \streetaddress{Oettingenstr. 67}
  \city{Munich}
  \state{Bavaria}
  \country{Germany}
  \postcode{80538}}
\email{jonas.nuesslein@ifi.lmu.de}

\author{Sebastian Zielinski}
\affiliation{%
  \institution{LMU Munich}
  \streetaddress{Oettingenstr. 67}
  \city{Munich}
  \state{Bavaria}
  \country{Germany}
  \postcode{80538}}
\email{sebastian.zielinski@ifi.lmu.de}

\author{Michael Kölle}
\affiliation{%
  \institution{LMU Munich}
  \streetaddress{Oettingenstr. 67}
  \city{Munich}
  \state{Bavaria}
  \country{Germany}
  \postcode{80538}}
\email{michael.koelle@ifi.lmu.de}

\author{Claudia Linnhoff-Popien}
\affiliation{%
  \institution{LMU Munich}
  \streetaddress{Oettingenstr. 67}
  \city{Munich}
  \state{Bavaria}
  \country{Germany}
  \postcode{80538}}
\email{linnhoff@ifi.lmu.de}

\renewcommand{\shortauthors}{Stein et al.}

\begin{abstract}
Quantum computing provides powerful algorithmic tools that have been shown to outperform established classical solvers in specific optimization tasks. A core step in solving optimization problems with known quantum algorithms such as the Quantum Approximate Optimization Algorithm (QAOA) is the problem formulation. While quantum optimization has historically centered around Quadratic Unconstrained Optimization (QUBO) problems, recent studies show, that many combinatorial problems such as the TSP can be solved more efficiently in their native Polynomial Unconstrained Optimization (PUBO) forms. As many optimization problems in practice also contain continuous variables, our contribution investigates the performance of the QAOA in solving continuous optimization problems when using PUBO and QUBO formulations. Our extensive evaluation on suitable benchmark functions, shows that PUBO formulations generally yield better results, while requiring less qubits. As the multi-qubit interactions needed for the PUBO variant have to be decomposed using the hardware gates available, i.e., currently single- and two-qubit gates, the circuit depth of the PUBO approach outscales its QUBO alternative roughly linearly in the order of the objective function. However, incorporating the planned addition of native multi-qubit gates such as the global Mølmer-Sørenson gate, our experiments indicate that PUBO outperforms QUBO for higher order continuous optimization problems in general.
\end{abstract}




\maketitle

\section{Introduction}
\label{sec:introduction}
Solving optimization problems is a central task in industries involving domains like production and logistics. Many of these problems concern scheduling, routing, packing and others, which are often NP-hard and thus demand for heuristic solvers. A particularly promising approach to solving such optimization problems is quantum computing, which has already shown results comparable to classical state-of-the-art methods for small problem sizes \cite{PhysRevX.6.031015, doi:10.1126/science.abo6587, PhysRevX.8.031016} despite current quantum hardware limitations. For a significant period of time, quantum optimization was driven by D-Wave System's Quantum Annealing devices, which are technically limited to solving problems written in Quadratic Unconstrained Binary Optimization (QUBO) form. This restriction was subsequently lifted in the Quantum Approximate Optimization Algorithm (QAOA) by Farhi et al., which essentially simulates the process of Quantum Annealing on a quantum gate computer and allows for additional generalization using the larger capabilities of a universal quantum computer \cite{farhi2014quantum}.

One particularly powerful generalization of the QAOA is its ability to solve higher order polynomial problems, i.e., it can natively work with Polynomial Unconstrained Binary Optimization (PUBO) problems. Instead of having to quadratize the a PUBO problem into QUBO form using ancillary qubits as is necessary for D-Wave's Quantum Annealers, needed multi-qubit interactions can be modelled using quantum gates \cite{nielsen_chuang_2010}. While current quantum computers generally only support single- and two-qubit gates, e.g., trapped ion quantum computers are expected to implement multi-qubit gates such as the (global) Mølmer-Sørenson gate in the future\footnote{\url{https://ionq.com/docs/getting-started-with-native-gates}}. Such gates will allow the execution of the qubit interactions necessary to model PUBO problems in constant time without the currently needed decomposition in two- and single-qubit gates \cite{Maslov_2018}, which scales linearly in the number of qubits involved.

While some binary, combinatorial optimization problems like Max-Cut or Number Partitioning are formulated in terms of QUBO natively, modelling intrinsically non-binary problems like the TSP for QUBO requires special encoding techniques like the \emph{one-hot encoding}, which increase the search space beyond exigence \cite{10.1007/s11128-021-03405-5}. For problems like these, it has been shown that their PUBO versions generally outperform their QUBO analogues in terms of solution quality as well as the required number of optimization steps and QAOA iterations \cite{10.1007/s11128-021-03405-5, 9259934}.

As many NP-hard problems such as scheduling or packing also involve continuous variables in higher order terms frequently in application \cite{Floudas2005}, we set out to compare the performance of PUBO and QUBO formulations for the QAOA on continuous optimization problems. Our two core contributions to this investigation are:
\begin{itemize}
    \item an implementation of the QAOA capable of solving arbitrary polynomial optimization problems, that allows control over the used bit depth and the domains of the input variables, and
    \item an in-depth case-study evaluating the performance of PUBO and QUBO problem formulations on two established, continuous optimization benchmark functions. 
\end{itemize}

This paper is structured into five sections. Following this introduction, we visit fundamental background knowledge necessary to comprehend our methodology in section \ref{sec:background}. Section \ref{sec:concept} subsequently contains a detailed description of the concept used to solve higher order continuous optimization problems with the QAOA. Finally, the established approach is applied to conduct the aspired evaluation in section \ref{sec:evaluation} while concluding with a contextualization of the acquired results in section \ref{sec:conclusion}.

\section{Background}
\label{sec:background}
In this section, we describe the overall functionality of the QAOA and its initial motivation to get an overview of all its components possibly influencing the evaluation results.

The QAOA is inspired by Adiabatic Quantum Computing (AQC), which is an alternative paradigm of quantum computing besides the omnipresent Quantum Gate Model (QGM). The main difference of AQC to the QGM resides in its time evolution being inherently continuous instead of iteratively applying discrete gates, as done in the QGM. Drawing upon the \emph{adiabatic theorem}, which essentially states that a physical system stays in its instantaneous eigenstate whenever the time evolution applied to it happens slowly enough and if there is a gap between the corresponding eigenvalue and the rest of the Hamiltonian's spectrum \cite{Born1928}, an optimization algorithm can be formulated as: 
\begin{enumerate}
    \item Prepare an initial state $\ket{\psi}$ that is the ground state of a known Hamiltonian $\hat{H}_M$.
    \item Identify a Hamiltonian $\hat{H}_C$ modelling the objective function $f:\left\lbrace 0,1\right\rbrace^{n}\rightarrow \mathbb{R}$ where the eigenstates represent possible solutions to the input problem. The eigenvalues that correspond to the eigenstates embody the objective values of the respective solution.
    \item \label{itm:annealing-process} Gradually evolve the initial state to the ground state of $\hat{H}_C$ corresponding to the global optimum of $f$ by applying the Hamiltonian $\hat{H}(t)=\left(1-t\right)\hat{H}_M + t\hat{H}_C$.
\end{enumerate}
The standard choice for the Hamiltonian $\hat{H}_M$ is $\hat{H}_M\coloneqq - \sum_{i=1}^{n} \sigma^x_i$ which inherits the easy to prepare ground state $\ket{+}^{\otimes n}$, where $\sigma^x_i$ denotes the tensor product of $n-1$ identity matrices $I$ with the Pauli operator $\sigma_x$ at the $i$-th position. For $\hat{H}_C$, a possible definition is $\hat{H}_C\coloneqq \sum_{x\in\left\lbrace 0,1\right\rbrace^{n}} f(x)\ket{x}\bra{x}$ as this trivially matches its requirements stated above.

While Quantum Annealers are built to execute the procedure described in item \ref{itm:annealing-process} for any given Ising Hamiltonian\footnote{Ising Hamiltonians represent the energy spectrum in a specific physical system. This system is described by an \emph{Ising model}, which is a mathematical model of ferromagnetism in statistical mechanics. This Hamiltonian has the convenient property of being isomorphic to the NP-hard quadratic programming problem and hence naturally allows to model many interesting optimization problems with it.} $\hat{H}_C=\sum_i h_i \sigma^z_i + \sum_{i<j}J_{ij}\sigma^z_i\sigma^z_j$, discretization and Hamiltonian simulation techniques must be used to implement this time evolution in the QGM, which is the fundamental idea of the QAOA. The continuous time evolution of $\hat{H}(t)$ is discretized by iteratively simulating the time evolution of the Hamiltonians $\hat{H}(t_k)$ with equidistant $t_k\in\left[0, 1\right]$ strictly increasing from $0$ to $1$ and $k\in\left\lbrace 1, ..., P\right\rbrace$.

To perfectly approximate the continuous time evolution in the limit for $P\rightarrow \infty$, each Hamiltonian $\hat{H}(t_k)$ is chosen to act for time $\nicefrac{1}{p}$. However, especially for small $P$, it is typically unclear how quickly the time evolution should progress at each intermediate Hamiltonian. In this context, it has proven useful to introduce parameters associated with the duration of their time evolution. These parameters can then be used to, i.a., satisfy the conditions of the adiabatic theorem, given that $P$ is big enough. Notably the concrete implementations proposed for this parameterization use independent parameters for both Hamiltonians: $\gamma_k\in\mathbb{R}$ for the Hamiltonian $\hat{H}_{C}$ and $\beta_k\in\mathbb{R}$ for the Hamiltonian $\hat{H}_{M}$. This allows for increased flexibility, especially in the regime of low $P$. For the optimization of these parameters, many different approaches have been explored, foremost gradient based techniques like the parameter shift rule in combination with gradient descent \cite{Mitarai_2018}, but also other heuristic approaches focused on yielding results very quickly, such as the COBYLA optimizer \cite{Powell1994}.

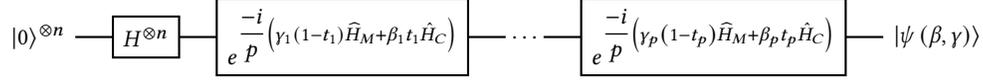
\begin{figure*}[h]
    \centering
\begin{quantikz}
\lstick{$\ket{0}^{\otimes n}$} & \gate{H^{\otimes n}} & \gate{e^{\dfrac{-i}{p} \left(\gamma_1 \left(1- t_1\right)\widehat{H}_{M} + \beta_1 t_1\hat{H}_{C}\right)}} & \qw \ \ldots \ & \gate{e^{\dfrac{-i}{p} \left(\gamma_p \left(1- t_p\right)\widehat{H}_{M} + \beta_p t_p\hat{H}_{C}\right)}} & \rstick{$\ket{\psi\left(\beta , \gamma\right)}$}\qw
\end{quantikz}
    \caption{The general form of the QAOA circuit. }
    \label{fig:QAOA-circ}
\end{figure*}

The QAOA algorithm can thus be understood as an algorithm, that simulates the time evolution of the Hamiltonian $\hat{H}(t)$ on gate based quantum computers. It does so using parameters guiding the time evolution speed as displayed in figure \ref{fig:QAOA-circ}.

\section{Concept}
\label{sec:concept}
In this section, we show how the QAOA can be used to solve higher order continuous polynomial optimization problems. More specifically, we employ the following procedures:
\begin{enumerate}
    \item Discretization of the objective function
    \item Translating the objective function into a Hamiltonian
    \item Implementing the Hamiltonian using quantum gates
\end{enumerate}

\subsection{Discretization of the objective function}\label{subsec:discretization}
For discretizing a given objective function $f:\left[a,b\right] \rightarrow \mathbb{R}$ with $a<b\in \mathbb{R}$, we need to select a suitable bit encoding. For the sake of simplicity, we choose the \emph{sign–magnitude} representation which maps any integer to its native binary encoding while initially disregarding its sign, to then finally represent its sign using an extra bit at the start, e.g.: $3_{10} \mapsto{} 0\, 11_2$ and $-3_{10} \mapsto{} 1\, 11_2$. In addition to that simplification, we also restrict the possible domain spaces of each variable to be of the form $\left]-2^n, 2^n\right[$ where $n\in\mathbb{N}$, to alleviate needed precautions for intervals that are unbalanced or away from powers of two. This decision allows us to incorporate numbers beyond the whole numbers in a straightforward manner, i.e., by using standard floating point representation with a freely selectable bit resolution $m\in\mathbb{N}$. The complete binary encoding of a given $x\in\left]-2^n, 2^n\right[$ and bit resolution $m\in\mathbb{N}$ can thus be described by the following approximation:
\begin{align}
    x \approx \left(2x_0-1\right)\left(\sum_{i=1}^{n} 2^{n-i} x_i + \sum_{i=1}^{m}x_{n+i}2^{-i}\right)
\end{align}
As desired, this discretization leads to the bit string representation $x\approx x_0 \, x_1 ... x_n,x_{n+1} ... x_{n+m}$, so that, e.g., $\left]-2^2, 2^2\right[\ni -2,75_{10} \mapsto{} 1\, 10,110_2$ for a bit resolution of $m=3$. Note however, that the borders of the domain space can only be approached when increasing the bit resolution $m$, while every additional bit contributes with advancement of $\nicefrac{1}{2^{m+1}}$. Using this bit encoding, we can also represent functions with higher dimensional input spaces by following the described substitution procedure for every dimension and then concatenating the resulting bit strings.

\subsection{Translating the objective function into a Hamiltonian}
\label{subsec:obj-fun-into-Ham}
As described in section \ref{sec:background}, there is a native mapping between binary functions $f:\left\lbrace 0,1\right\rbrace^{n}\rightarrow \mathbb{R}$ and Hamiltonians, i.e., $\hat{H}_C\coloneqq \sum_{x\in\left\lbrace 0,1\right\rbrace^{n}} f(x)\ket{x}\bra{x}$. This method can be very inefficient however, if we only have access to $f$ as a black box function, because the Hamiltonian can be comprised of exponentially many non-zero terms. Given that we have access to $f$ in a white box manner, we can conduct this mapping much more efficiently, i.e., by substituting every $x_i\in\left\lbrace 0,1\right\rbrace$ with a $s_i\in\left\lbrace -1,1\right\rbrace$ as in $x_i\mapsto\nicefrac{\left(s_i+1\right)}{2}$. In the case of $f$ having higher degree interactions than two in its input bits (i.e., e.g., a term like $\alpha x_0x_1x_2$ with $\alpha\in\mathbb{R}$), inserting a suitable quadratization step is obligatory for the QUBO version. Typically this step is done before translating into the spin configuration domain $\left\lbrace -1,1\right\rbrace$ by adding ancillary bits to the input space and a penalty term to the function $f$, as exemplified in equation \ref{eq:quadratization}. For details on this quadratization step, we reference to the python package qubovert, which we used for this step in our implementation\footnote{\url{https://github.com/jtiosue/qubovert}}. Notably, finding the optimal quadratization in terms of minimizing the number of needed ancillary qubits is NP-hard, as pointed out in \cite{BOROS2002155}.
\begin{align}
    f(x_0,x_1,x_2)&=\alpha x_0x_1x_2 \nonumber\\
    \mapsto\; f(x_0,x_1,x_2,z)&=\alpha zx_2+2\alpha \left(x_0x_1-2\left(x_0+x_1\right)z+3z\right)
    \label{eq:quadratization}
\end{align}
In order to translate the resulting function of spin configurations $f':\left\lbrace -1,1\right\rbrace^{n}\rightarrow \mathbb{R}$ into a quantum mechanical Hamiltonian, we can simply substitute all spins $s_i$ with Pauli operators using the trivial map $s_i\mapsto\sigma_{i}^z$. \cite{farhi2014quantum}

\subsection{Implementing the Hamiltonian using quantum gates}
\begin{figure*}[ht!]
\centering
\begin{subfigure}[b]{0.3\textwidth}    
        \centering
        \begin{quantikz}
\qw & \gate{R_z\left(\theta\right)} & \qw 
\end{quantikz}
\caption{Hamiltonian simulation of $\dfrac{\theta}{2}\sigma_{0}^z\sigma_{1}^z$.}
        \label{fig:Ham-Sim-1D}
    \end{subfigure}
    \hfill
\begin{subfigure}[b]{0.3\textwidth}    
        \centering
        \begin{quantikz}
\qw & \ctrl{1} & \qw & \ctrl{1} &   \qw \\
\qw & \targ{} & \gate{R_z\left(\theta\right)} & \targ{} &  \qw 
\end{quantikz}
\caption{Hamiltonian simulation of $\dfrac{\theta}{2}\sigma_{0}^z\sigma_{1}^z$.}
        \label{fig:Ham-Sim-2D}
    \end{subfigure}
    \hfill
    \begin{subfigure}[b]{0.3\textwidth}
        \centering
        \begin{quantikz}
\qw & \ctrl{1} & \qw & \qw & \qw & \ctrl{1} &\qw \\
\qw & \targ{} & \ctrl{1} & \qw & \ctrl{1} &  \targ{} &  \qw \\
\qw & \qw & \targ{} & \gate{R_z\left(\theta\right)} & \targ{} &  \qw &  \qw 
\end{quantikz}
        \caption{Hamiltonian simulation of $\dfrac{\theta}{2}\sigma_{0}^z\sigma_{1}^z\sigma_{2}^z$.}
\label{fig:Ham-Sim-3D}
    \end{subfigure}
    \caption{Hamiltonian simulation of the components in the cost Hamlitonian $\hat{H}_C$.}
    \label{fig:Ham-Sim-Pali-Tensors}
\end{figure*}
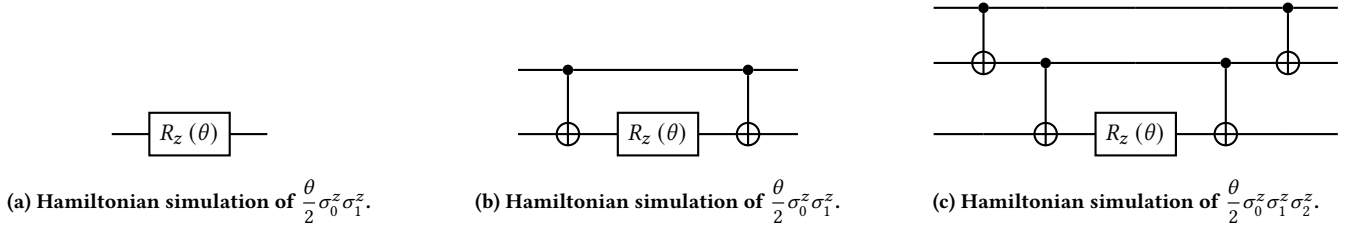

To implement the quantum circuit of the QAOA, we have to conduct Hamiltonian simulation of $\hat{H}_M$ and $\hat{H}_C$. While $\hat{H}_M$ can easily be simulated using parameterized $X$ gates, $\hat{H}_C$ involves higher order terms (as e.g., $\alpha \sigma_{i}^z \sigma_{j}^z \sigma_{k}^z $ where $\alpha \in \mathbb{R}$) for the PUBO variant. As pointed out in \cite{Glos2022}, Hamiltonians of this form can be simulated using the generic architecture shown in figure \ref{fig:Ham-Sim-Pali-Tensors}, naturally expanding from the well-know quadratic case $\alpha \sigma_{i}^z \sigma_{j}^z$. When having access to a suitable multi-qubit gate such as the (global) Mølmer-Sørenson gate, combining the information presented in figure 4.19 in \cite{nielsen_chuang_2010} and figure 5 from \cite{Maslov_2018}, we can simulate arbitrary degrees of Pauli matrices using one extra ancillary qubit with an overhead of merely two extra circuit operations. As all terms in $\hat{H}_C$ commute, the Hamiltonian simulation simplifies into a concatenation of the gates used to implement all terms in the sum notation of $\hat{H}_C$ as exemplified in figure \ref{fig:puboCostHamilCircuit}, concluding this section.

\subsection{Example}\label{subsec:example}
\begin{figure*}[h]
    \centering
\begin{quantikz}[column sep=12pt, row sep={20pt,between origins}]
\lstick{$\ket{0}$} & \gate{H}\gategroup[3,steps=1,style={dashed, rounded corners,fill=blue!20, inner xsep=2pt}, background,label style={label position=below,anchor= north,yshift=-0.2cm}]{{ State prep.}} & \gate{R_z\left(16\gamma_1\right)}\gategroup[3,steps=10,style={dashed, rounded corners,fill=blue!20, inner xsep=2pt}, background,label style={label position=above,anchor= north,yshift=0.5cm}]{{Parameterized Hamiltonian simulation of $\hat{H}_C$}} & \ctrl{1} & \qw & \ctrl{1} & \ctrl{2} & \qw & \ctrl{2} & \qw & \qw & \qw & \gate{R_x(2\beta_1)}\gategroup[3,steps=1,style={dashed, rounded corners,fill=blue!20, inner xsep=2pt}, background,label style={label position=below,anchor= north,yshift=-0.2cm}]{{Parameterized Hamiltonian simulation of $\hat{H}_M$}} & \qw \ \ldots \ & \meter{}\qw \\
\lstick{$\ket{0}$} & \gate{H} & \gate{R_z\left(16\gamma_1\right)} & \targ{} & \gate{R_z\left(8\gamma_1\right)} & \targ{} & \qw & \qw & \qw  & \ctrl{1} & \qw & \ctrl{1} &  \gate{R_x(2\beta_1)} & \qw \ \ldots \ & \meter{}\qw \\
\lstick{$\ket{0}$} & \gate{H} & \gate{R_z\left(16\gamma_1\right)} & \qw & \qw & \qw & \targ{} & \gate{R_z\left(8\gamma_1\right)} & \targ{} & \targ{} & \gate{R_z\left(8\gamma_1\right)} & \targ{} &\gate{R_x(2\beta_1)} & \qw \ \ldots \  & \meter{}\qw
\end{quantikz}
    \caption{QAOA circuit implementation using single-qubit and CNOT-gates for the example in section \ref{subsec:example} showing $P = 1$ iterations.}
    \label{fig:puboCostHamilCircuit}
\end{figure*}
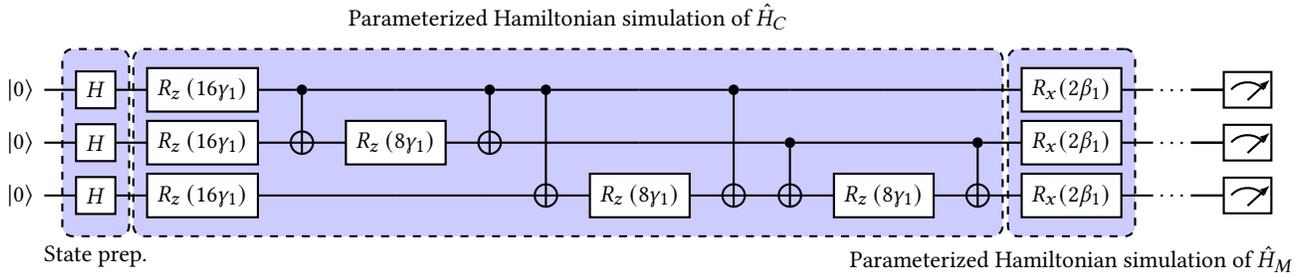

We now demonstrate how all described steps of transforming the objective function into the corresponding QAOA circuit can be done in practice using the following example:
\begin{align}
    f:\left]-2^2,2^2\right[&\rightarrow \mathbb{R}\\
    x&\mapsto x^2 + 2x
\end{align}
Choosing a zero bit resolution $m=0$ for simplicity, the bit encoding is displayed in the following map:
\begin{align}
    x\mapsto \left(2x_0-1\right)\left(2^1 x_1 + 2^0 x_2\right).
\end{align}
Therefore, $f$ can now be written in discretized form as follows:
\begin{align}
    f\left(x_0,x_1,x_2\right)= &\left(\left(2x_0-1\right)\left(2^1 x_1 + 2^0 x_2\right)\right)^2\nonumber \\
    &+ 2\left(2x_0-1\right)\left(2^1 x_1 + 2^0 x_2\right) \nonumber \\
    =& 4\left(4x_0x_1+x_0x_2+x_1x_2\right)
\end{align}
This then translates to the spin configuration function $f'$ as described in equation \ref{eq:spin-conf-fun} below.
\begin{align}
    f'\left(s_0,s_1,s_2\right)=& 4\left(4\dfrac{s_0+1}{2}\dfrac{s_1+1}{2}+\dfrac{s_0+1}{2}\dfrac{s_2+1}{2}+\dfrac{s_1+1}{2}\dfrac{s_2+1}{2}\right) \nonumber \\
    =&4\left(s_0s_1+s_0s_2+s_1s_2+2s_0+2s_1+2s_2+3\right)
    \label{eq:spin-conf-fun}
\end{align}
Using the mapping from a spin configuration function to a quantum Hamiltonian as described in section \ref{subsec:obj-fun-into-Ham}, we get:
\begin{align}
        \hat{H}_C=4\left(\sigma_{0}^z\sigma_{1}^z+\sigma_{0}^z\sigma_{2}^z+\sigma_{1}^z\sigma_{2}^z+2\sigma_{0}^z+2\sigma_{1}^z+2\sigma_{2}^z+3I^{\otimes 3}\right)
\end{align}
Subsequently, we can use the combination of CNOT gates wrapping a parameterized rotation gate $R_z(\theta)$ applied on the target qubit to construct the circuit simulating the Hamiltonian $\hat{H}_C$, as indicated figure \ref{fig:puboCostHamilCircuit}.

\section{Evaluation}
\label{sec:evaluation}
\begin{figure*}[h!]
\centering
\begin{subfigure}[b]{0.45\textwidth}    
        \centering
        \includegraphics[width=\textwidth]{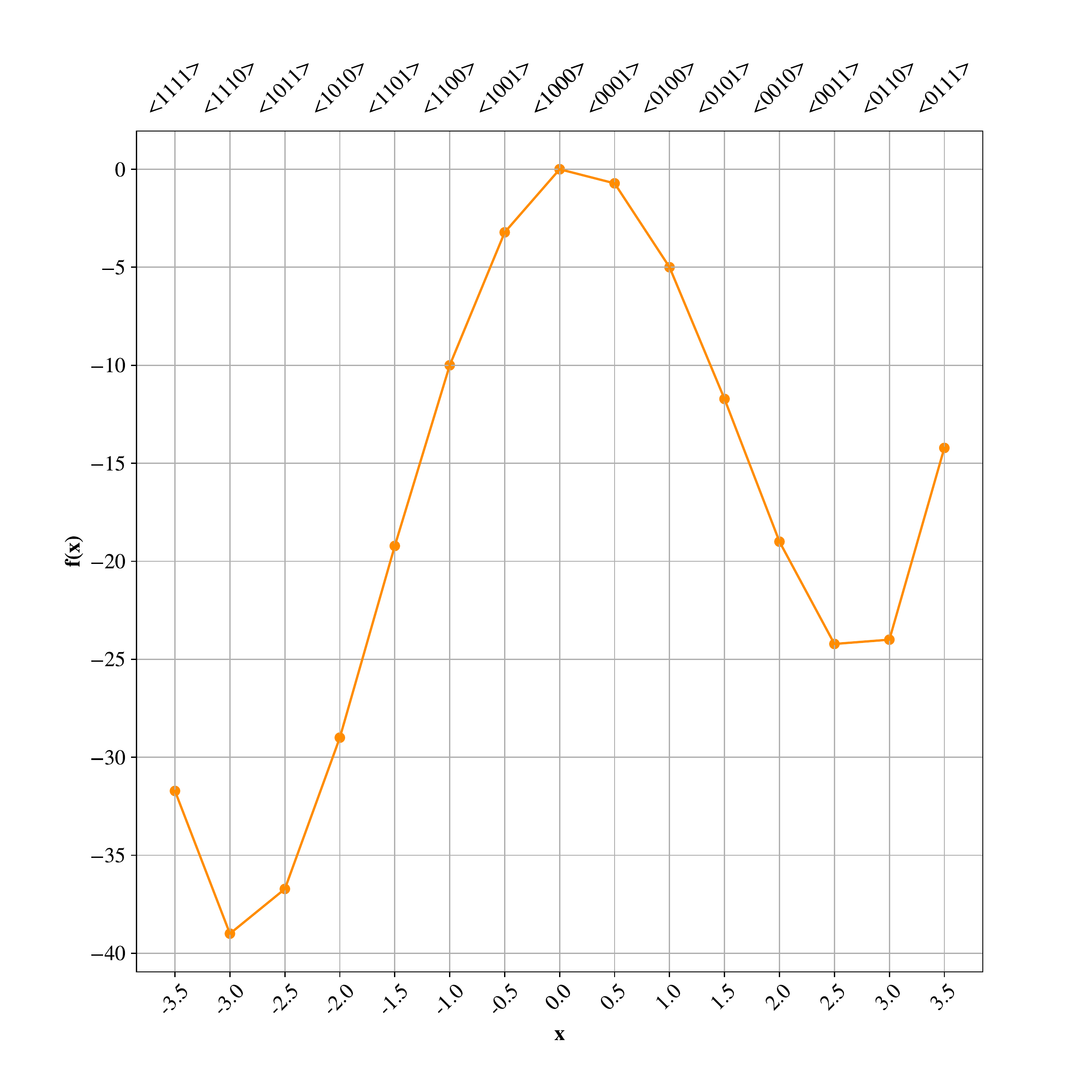}
\caption{The 1D Styblinski-Tang function for a bit resolution of one.}
        \label{fig:DiscSTangBitRes1}
    \end{subfigure}
    \hfill
    \begin{subfigure}[b]{0.45\textwidth}
        \centering
        \includegraphics[width=\textwidth]{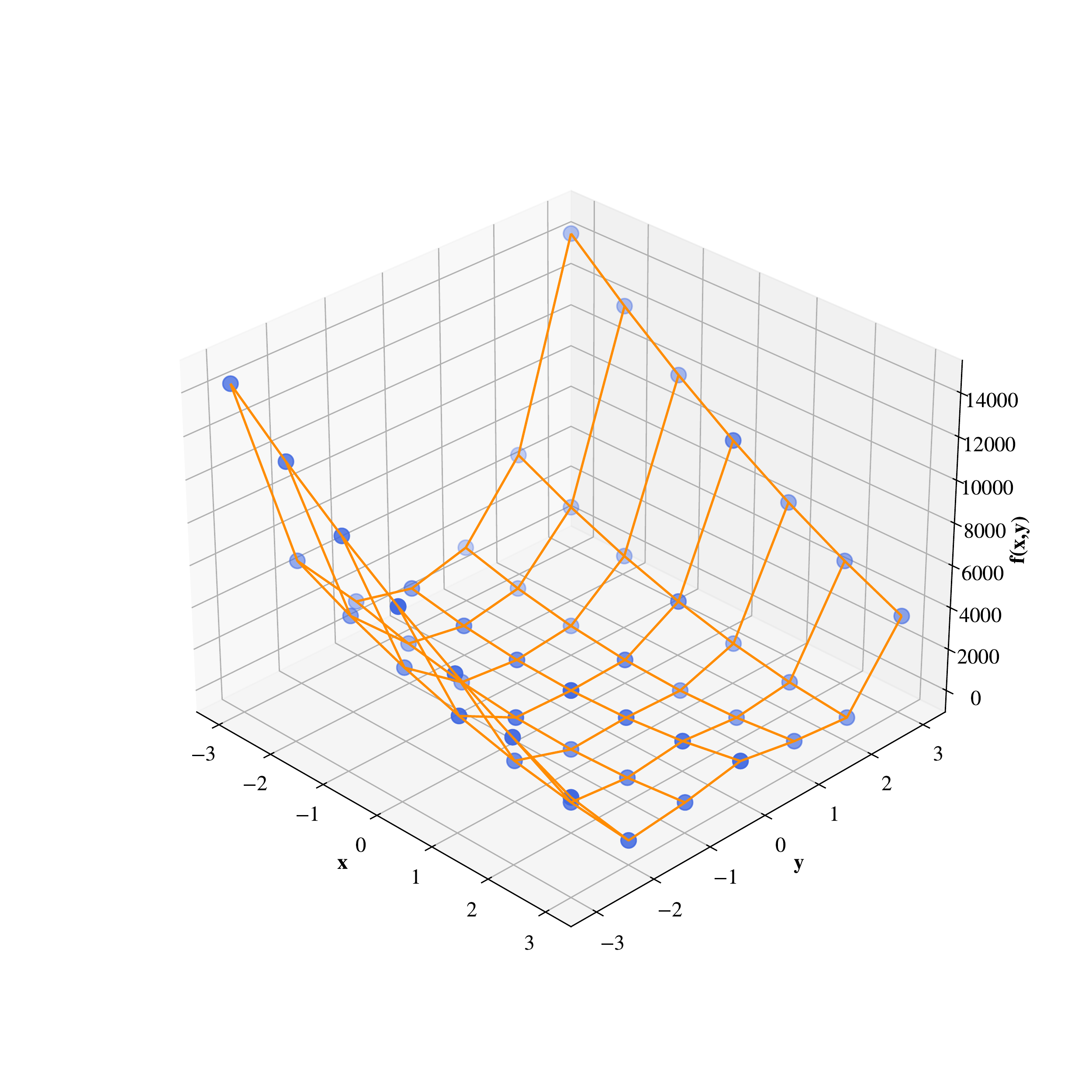}
        \caption{The 2D Rosenbrock function for a bit resolution of zero.}
\label{fig:DiscRosenBitRes0}
    \end{subfigure}
    \label{fig:benchmarkfcts}
    \caption{Visualizations of benchmark functions used for the evaluation.}
\end{figure*}

To compare the performance of the QAOA for PUBO and QUBO formulations of higher order continuous optimization functions, we run experiments on two established benchmark functions (see figures \ref{fig:DiscSTangBitRes1} and \ref{fig:DiscRosenBitRes0}): The 1-Dimensional Styblinski-Tang function $s(x)=\nicefrac{\left(x^4-16x^2+5x\right)}{2}$ (denoted as 1D-ST) \cite{STYBLINSKI1990467}, and the 2-Dimensional Rosenbrock function $r(x,y)=100\left(y-x^2\right)^2 +\left(x-1\right)^2$ (denoted as 2D-Rb) \cite{10.1093/comjnl/3.3.175}. These functions where chosen for their different requirements in terms of the number of needed qubits to model them (for details see figure \ref{fig:Circuit-Width-and-Depth}) and their hardness\footnote{According to the results from \emph{Global Optimization Benchmarks and AMPGO} by Andrea Gavana, see \url{http://infinity77.net/global_optimization/index.html}}. Having to specify input domain spaces in which the search for the optimal value is to be conducted, we choose the interval $\left]-4,4\right[$ for the 1D-ST function and $\left]-4,4\right[^2$ for the 2D-Rb function. These domain spaces allow us to find the  global optimum of each function and enable us to investigate many different bit resolutions while staying within reasonable simulation times of a couple of hours. More specifically, these input domains allow exploring bit resolutions of 0 to 3 for the 1D-ST function and 0 to 1 for the 2D-Rb function. 

In the following, we explore the performance differences between the PUBO and QUBO approaches in terms of three criteria:
\begin{enumerate}
    \item The solution quality
    \item The parameter training
    \item The circuit width and depth
\end{enumerate}
For all of the following experiments, we used Qiskit's qasm simulator, the COBYLA optimizer because of its short runtime, and 1024 shots as a standard for all circuit runs. In addition to that, we initialized all parameters using \emph{ramp initialization}, as it consistently showed the best results in our experiments. Notably, the ramp initialization simply corresponds to the choosing equidistantly spaced intervals for the discretized Hamiltonian simulation described in section \ref{sec:background}. Furthermore, we conducted our studies for a very high number of QAOA iterations compared to related work, i.e., $1\leq P \leq 40$, as this allows for a better performance estimation in terms of scaling.

\subsection{Solution quality}\label{EvalCritExpects}
To evaluate the solution quality of both approaches (PUBO and QUBO), we now examine their performance at different bit resolutions and varying QAOA iterations $P$ as exemplified in figure \ref{fig:boxplot-expval}.

For figure \ref{fig:ExpectSTangBitRes0M1}, we chose to display a baseline result, i.e., the 1D-ST function at zero bit resolution, as this function is a QUBO problem by nature. With both plots showing very similar behavior, it becomes apparent that PUBO performs completely analogously to the QUBO for quadratic functions.

Examining figures \ref{fig:ExpectSTangBitRes1M1} and \ref{fig:ExpectRosenBitRes0M1}, we can see that the PUBO approach consistently outperforms the QUBO approach for higher order functions, as the expected value, the median and the overall variance are significantly lower for PUBO. This becomes increasingly apparent for the harder 2D-Rb function displayed in figure \ref{fig:ExpectRosenBitRes0M1}, as we can see that the QUBO approach essentially plateaus for increasing $P$, while the PUBO performance clearly benefits from higher $P$. These results are especially promising if this trend does continue for higher $P$, which is to be explored in future work.

While these plots merely display exemplified results, our full evaluation results clearly substantiate the trends visible in the selected plots.

\newcommand{\lenExpectGraphs}{0.88\textwidth}
\begin{figure*}[ht]
    \begin{subfigure}[b]{\textwidth}    
        \centering
        \includegraphics[width=\lenExpectGraphs]{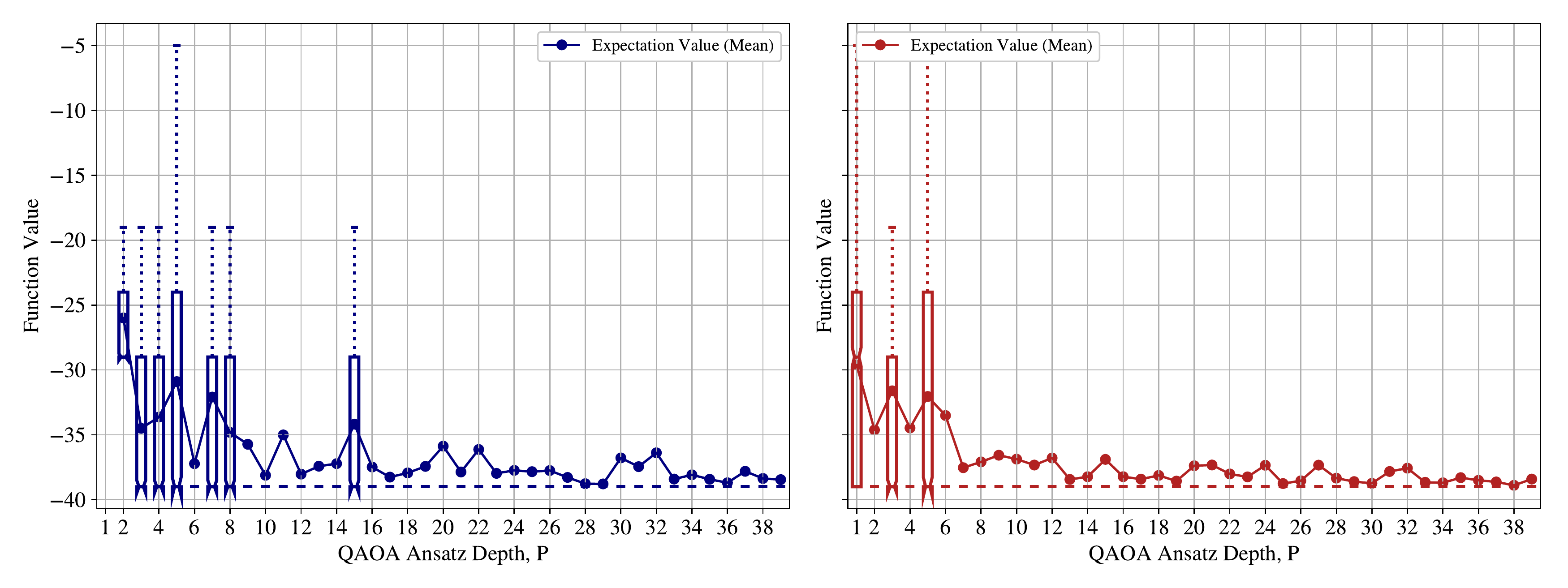}
        \caption{Results for the 1D-ST function with a bit resolution of zero.}
        \label{fig:ExpectSTangBitRes0M1}
    \end{subfigure}
    \hfill
    \begin{subfigure}[b]{\textwidth}
        \centering
        \includegraphics[width=\lenExpectGraphs]{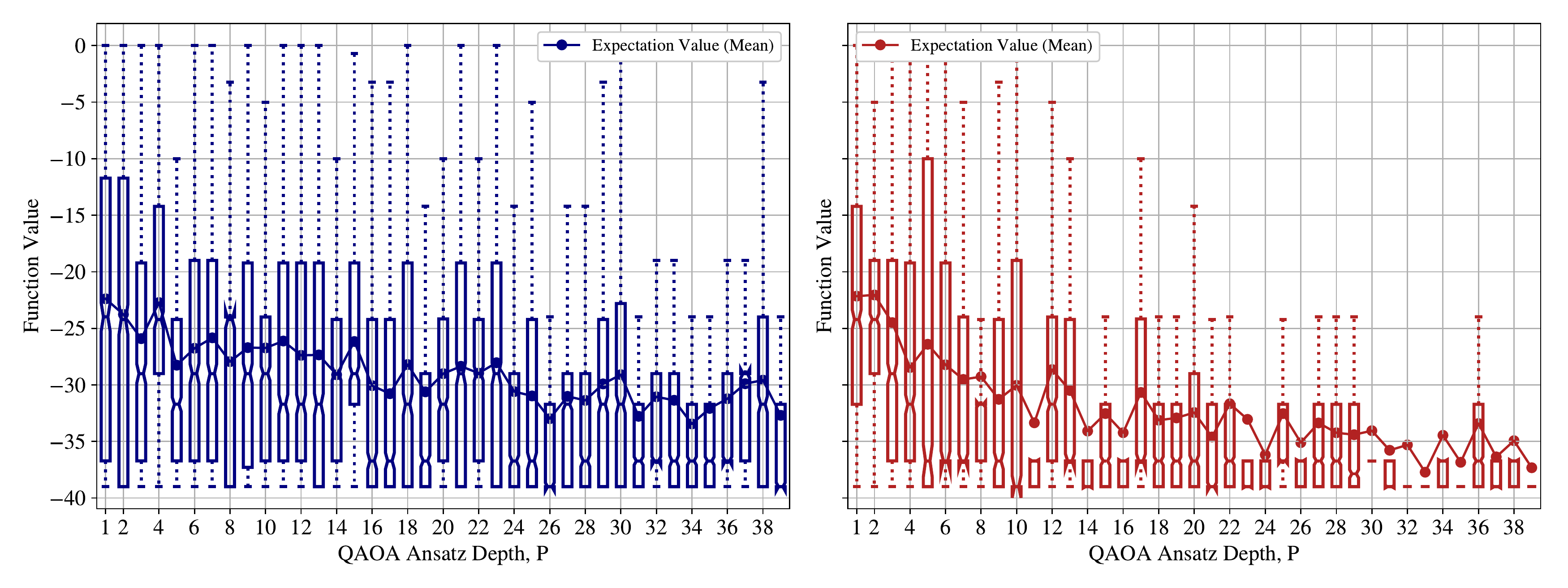}
        \caption{Results for the 1D-ST function with a bit resolution of one.}
        \label{fig:ExpectSTangBitRes1M1}
    \end{subfigure}
    \hfill
    \begin{subfigure}[b]{\textwidth}
        \centering
        \includegraphics[width=\lenExpectGraphs]{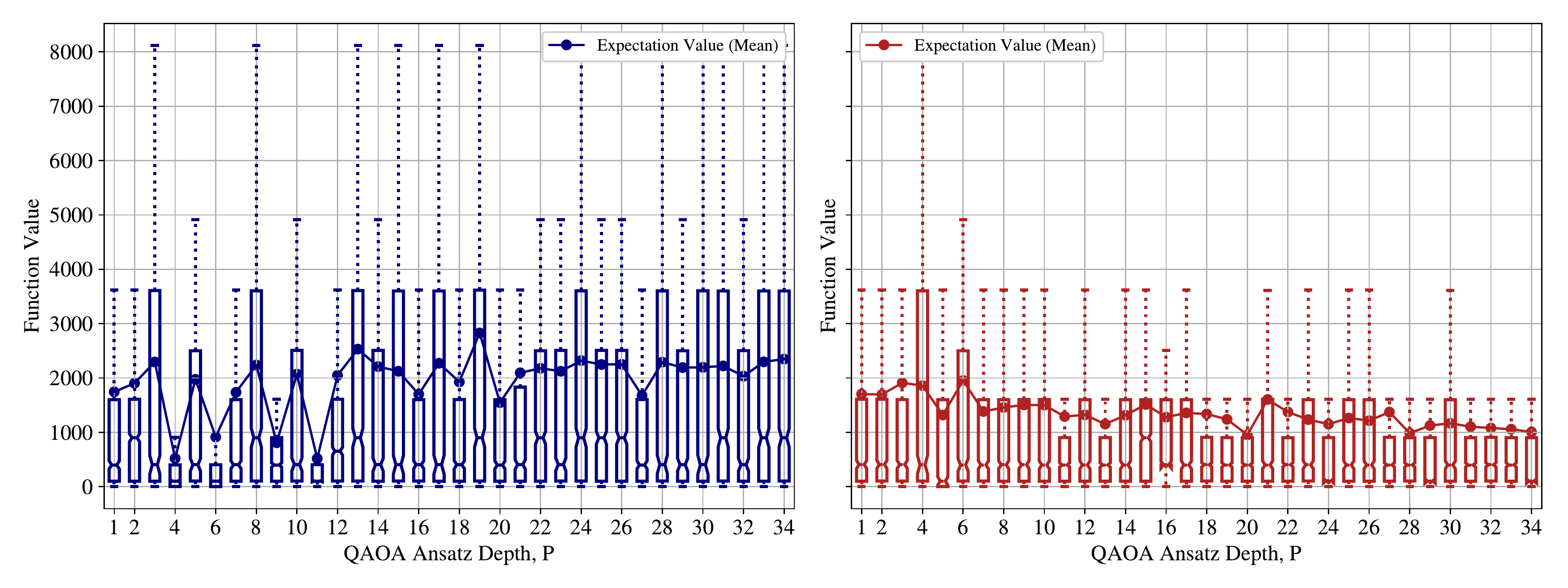}
        \caption{Results for the 2D-Rb with a bit resolution of zero.}
        \label{fig:ExpectRosenBitRes0M1}
    \end{subfigure}
    \caption{Box plots showing the quality of the solutions found using the QAOA for the QUBO (blue) and PUBO (red) approaches for different numbers of QAOA iterations $P$. The seeked global minimum for the Rosenbrock function is $0$ and $-39.16599$ for the Styblinksi-Tang.}
    \label{fig:boxplot-expval}
\end{figure*}

\subsection{Parameter Training}\label{EvalCritOptimTime}
Following the recommendation of \cite{team_2022}, we select the COBYLA optimizer to train the QAOA parameters. This optimizer has a built in stopping criterion, terminating the learning process when the last couple optimization iterations did not increase the objective value above a specific threshold (in our case $1e-4$). To prevent this procedure from exceeding a reasonable execution time, the user can also specify a number of maximum possible iterations. We use this functionality by capping the number of optimization steps at 1000, relying on results from preliminary experiments that showed, that almost no problem instances exceeded this number of optimization iterations. This allows us to compare the number of optimization steps between the PUBO and QUBO approaches unimpaired of this hyperparameter, as almost all parameter trainings run until completion.

Examining the number of optimization steps for different $P$ shown in figure \ref{fig:opt-iter}, it becomes clear that both approaches need roughly the same number of optimization steps. In general, we can also observe that the number of optimization steps for the Styblinski-Tang function is generally higher compared to the Rosenbrock function. We suspect this being caused by the flatter landscape of the Rosenbrock function leading to below-threshold training improvements sooner. In addition to that, we can observe that the QUBO approach has a tendency to decrease its ascend in training time earlier when the solution quality is worse than the PUBO (which is the case for the 1D-ST function at a bit resolution of three, as this function has a very similar plot to the one displayed in figure \ref{fig:ItrSTangBitRes3M1}).

\newcommand{\lenThreeGraph}{0.33}
\begin{figure*}[ht]
    \centering
    \begin{subfigure}[b]{\lenThreeGraph\textwidth}
        \includegraphics[width=\textwidth]{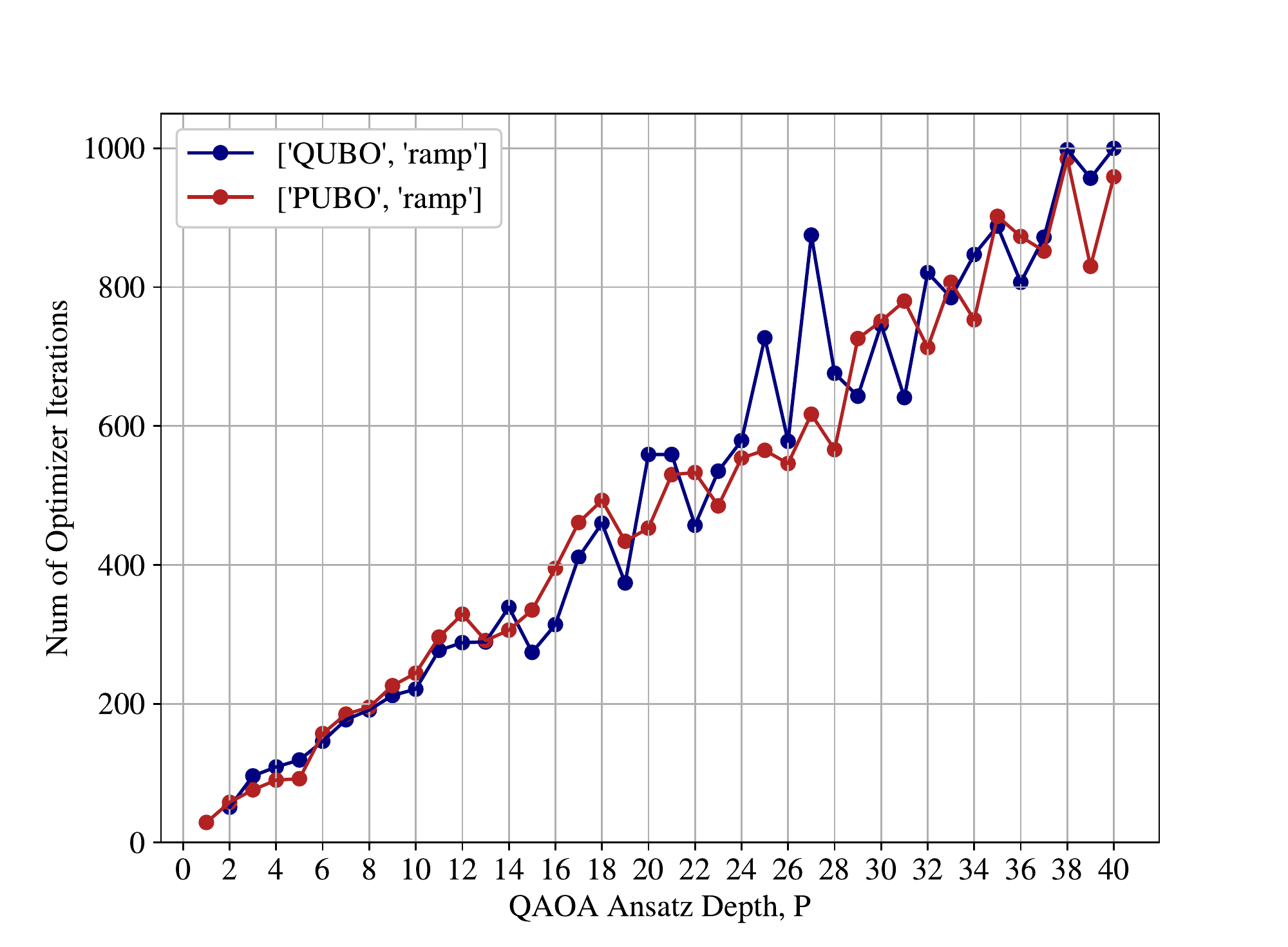}
        \caption{1D-ST function at a bit resolution of zero.}
        \label{fig:ItrSTangBitRes0M1}
    \end{subfigure}
    \hfill
    \begin{subfigure}[b]{\lenThreeGraph\textwidth}
        \centering
        \includegraphics[width=\textwidth]{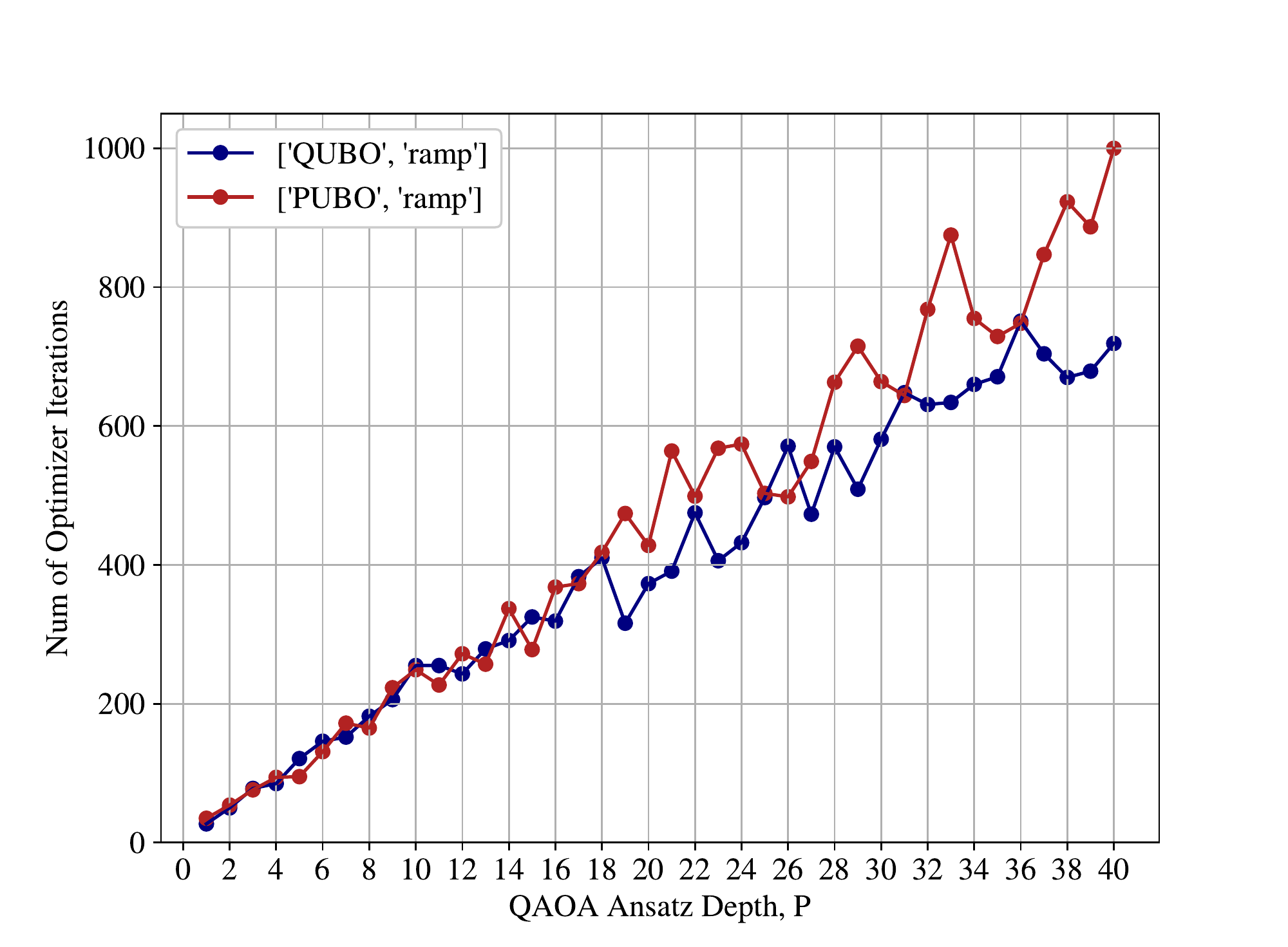}
        \caption{1D-ST function at a bit resolution of three.}
        \label{fig:ItrSTangBitRes3M1}
    \end{subfigure}
    \hfill
    \begin{subfigure}[b]{\lenThreeGraph\textwidth}
        \centering
        \includegraphics[width=\textwidth]{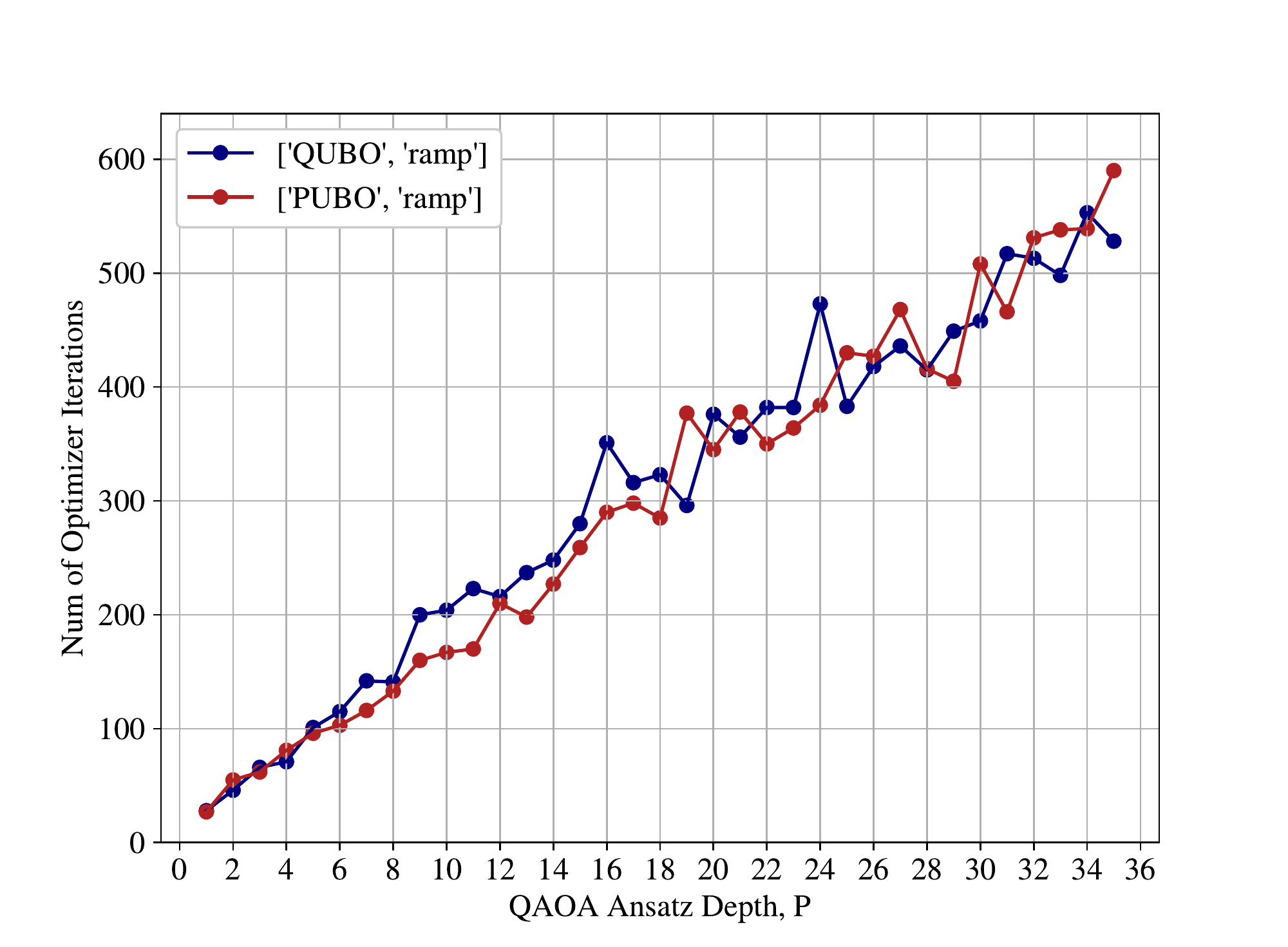}
        \caption{2D-Rb function at a bit resolution of one.}
        \label{fig:ItrRosenBitRes1M1}
    \end{subfigure}
    \caption{Number of parameter training iterations for different numbers of QAOA iterations $P$.}
    \label{fig:opt-iter}
\end{figure*}

\begin{figure*}[ht]
    \begin{subfigure}[b]{\lenThreeGraph\textwidth}
        \centering
        \includegraphics[width=\textwidth]{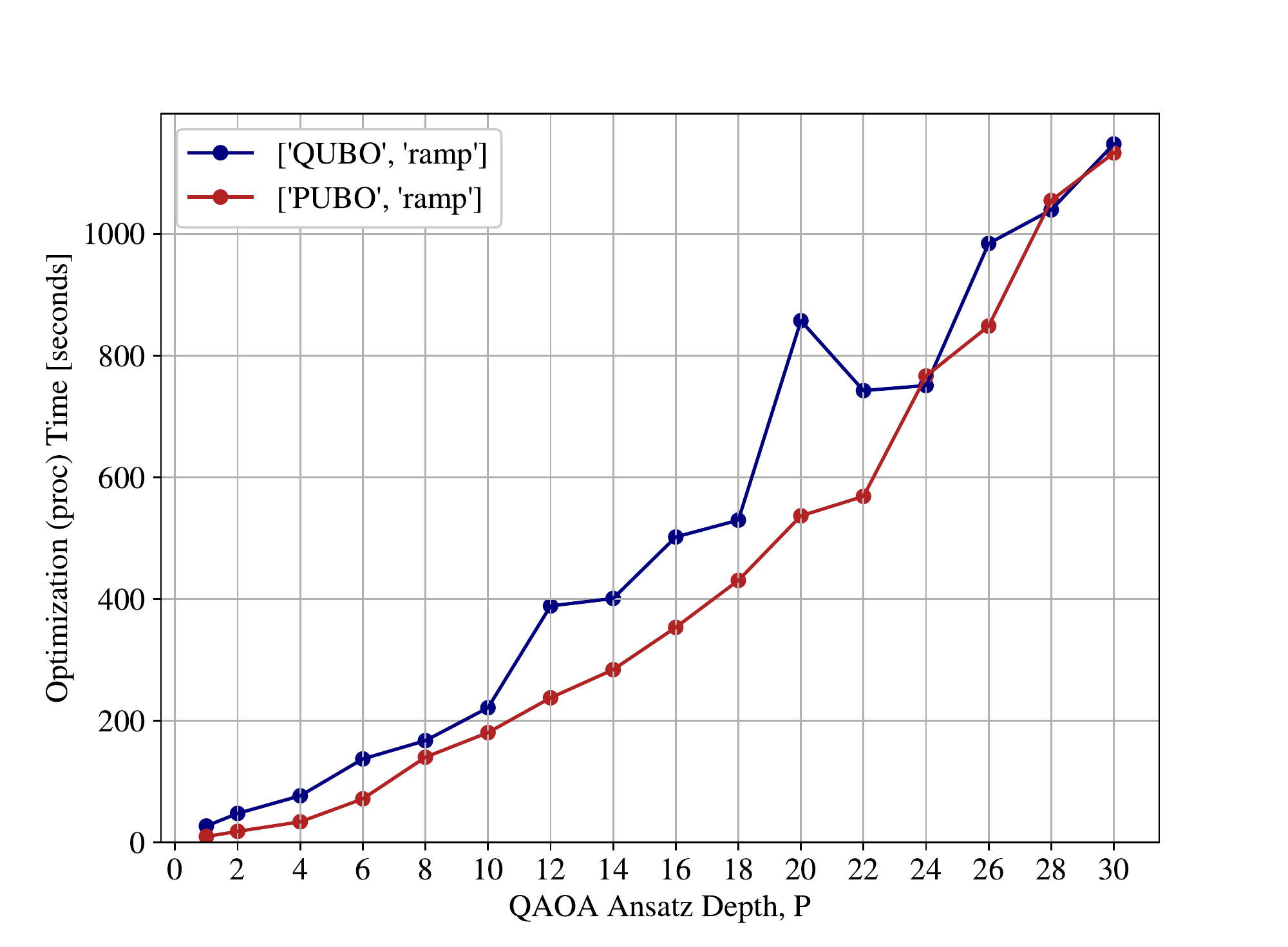}
        \caption{2D-Rb function at a bit resolution of zero.}
        \label{fig:OptimTimeRosenBitRes0M2}
    \end{subfigure}
    \hfill
    \begin{subfigure}[b]{\lenThreeGraph\textwidth}
        \centering
        \includegraphics[width=0.96\linewidth]{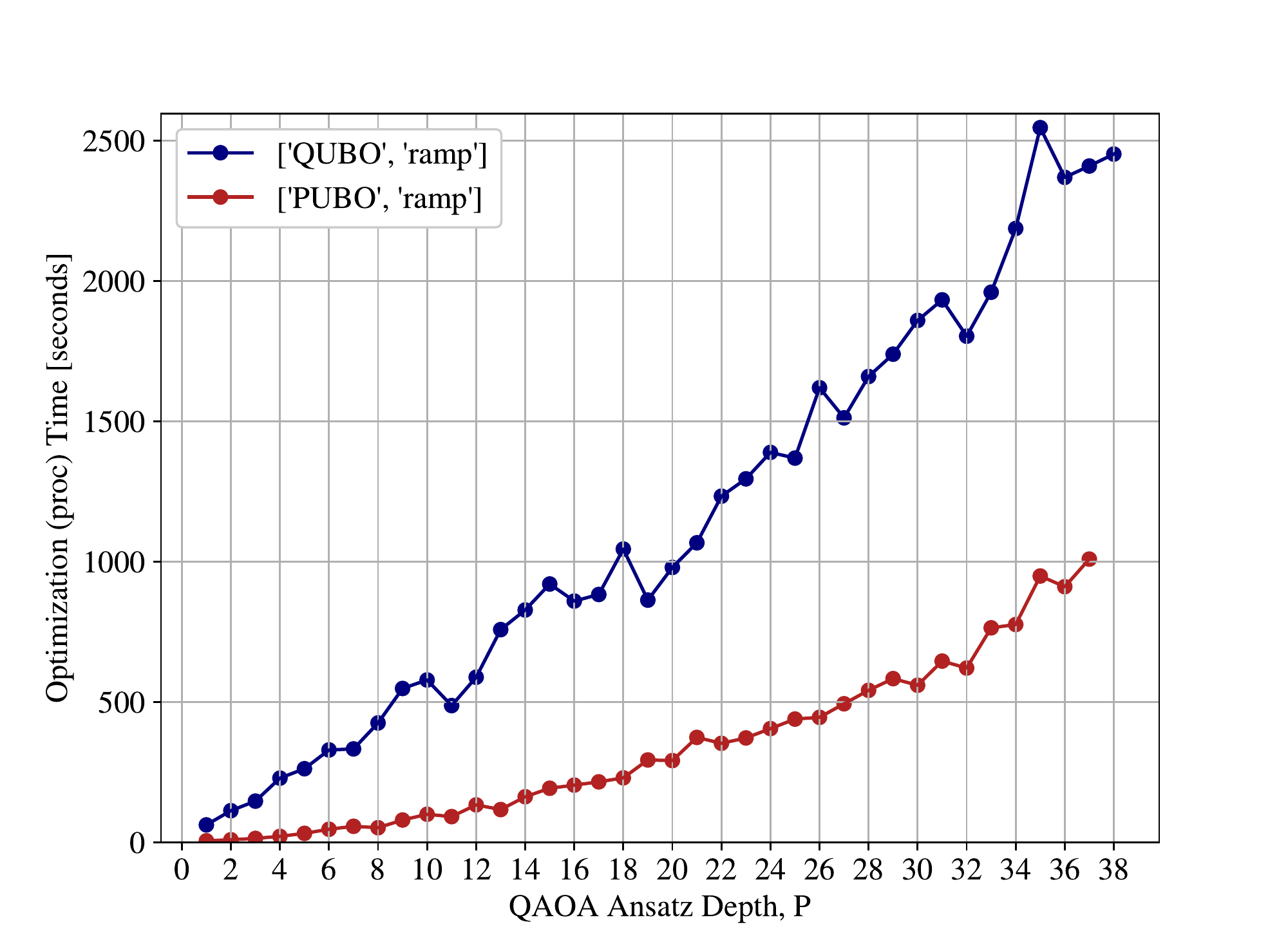}
        \caption{1D-ST, at a bit resolution of three.}
        \label{fig:OptimTimeSTangBitRes3M2}
    \end{subfigure}
    \hfill
    \begin{subfigure}[b]{\lenThreeGraph\textwidth}
        \centering
        \includegraphics[width=\textwidth]{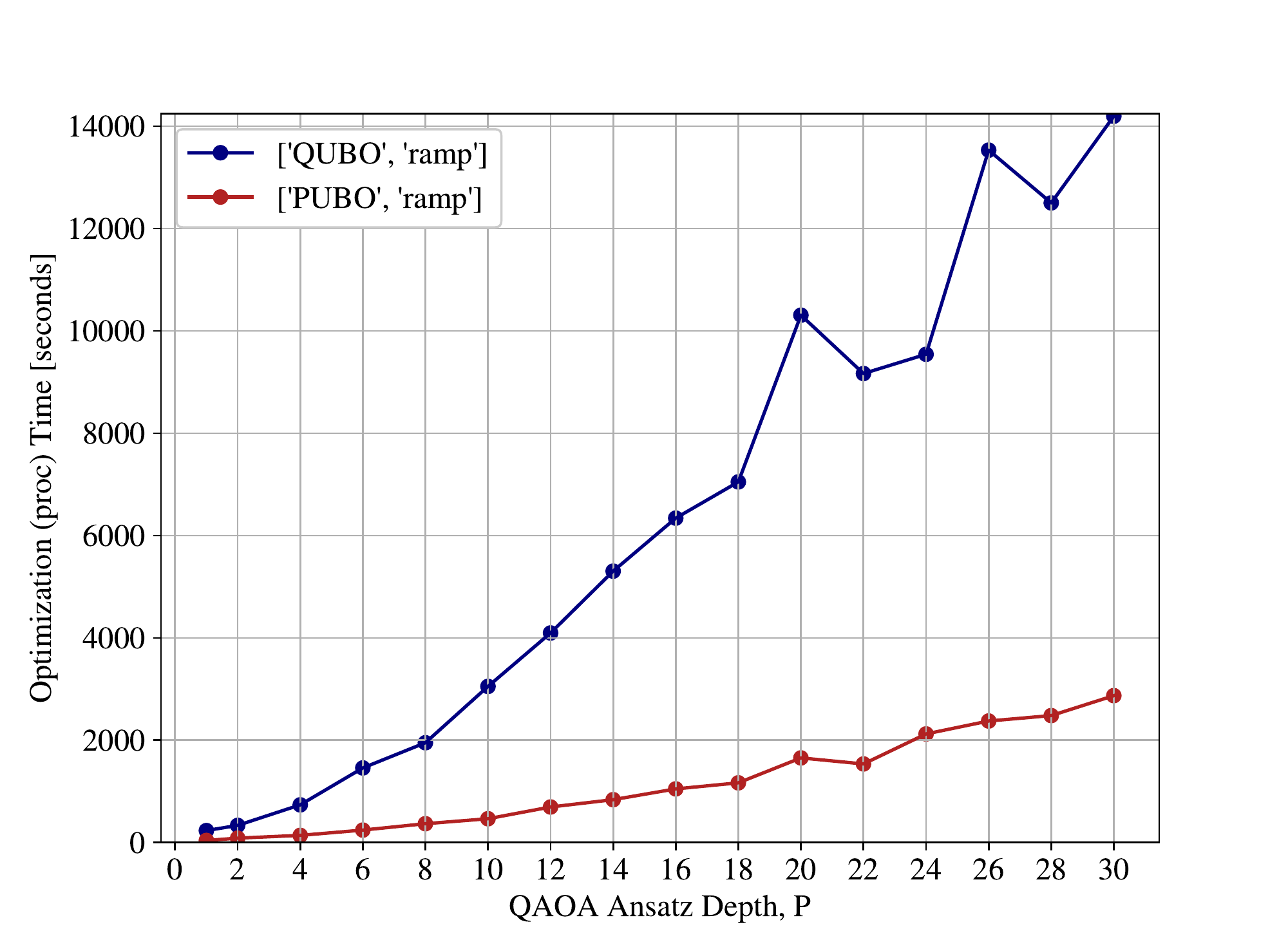}
        \caption{2D-Rb, at a bit resolution of one.}
        \label{fig:OptimTimeRosenBitRes1M2}
    \end{subfigure}
    \caption{Wall-clock training time of the QAOA parameters for different numbers of QAOA iterations $P$.}
    \label{fig:wallclock-training}
\end{figure*}

When simulating quantum circuits using classical hardware, execution times play an important role, as they limit what can be learned about their properties such as scaling behavior using non-quantum hardware. As displayed in figure \ref{fig:wallclock-training}, the training time does not differ significantly if the function only has a small amount of higher order terms involved (see figure \ref{fig:OptimTimeRosenBitRes0M2}), while execution time increases massively for the QUBO approach the more qubits are needed and the more higher order terms appear. Notably, that difference is mostly dominated by the number of qubits involved (13 for the QUBO formulation of 1D-ST at a bit resolution of three versus the 17 qubits needed for the QUBO formulation of the 2D-Rb function at a bit resolution of 1). This clearly demonstrates the performance of PUBO for simulation on classical hardware, also allowing for a deeper scaling analyses, which are very valuable in practice.

\subsection{Circuit width and depth}\label{EvalCritDepthVQ}
\begin{figure*}
    \begin{subfigure}[b]{0.4\textwidth}
        \centering
        \includegraphics[width=\textwidth]{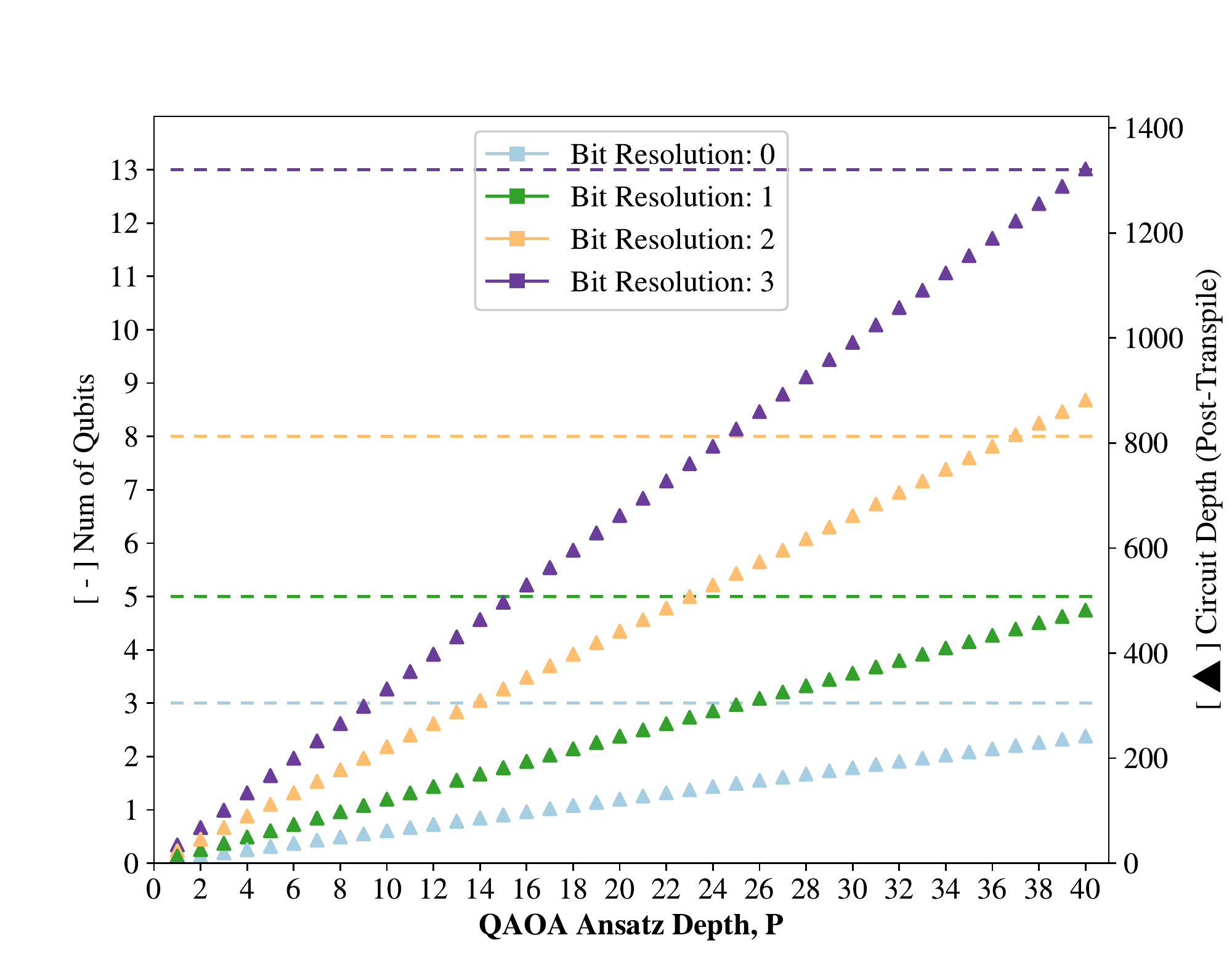}
        \caption{1D-ST, QUBO}
        \label{fig:QbitVDepthSTangQuboM1}
    \end{subfigure}
    \hfill
    \begin{subfigure}[b]{0.4\textwidth}
        \centering
        \includegraphics[width=\textwidth]{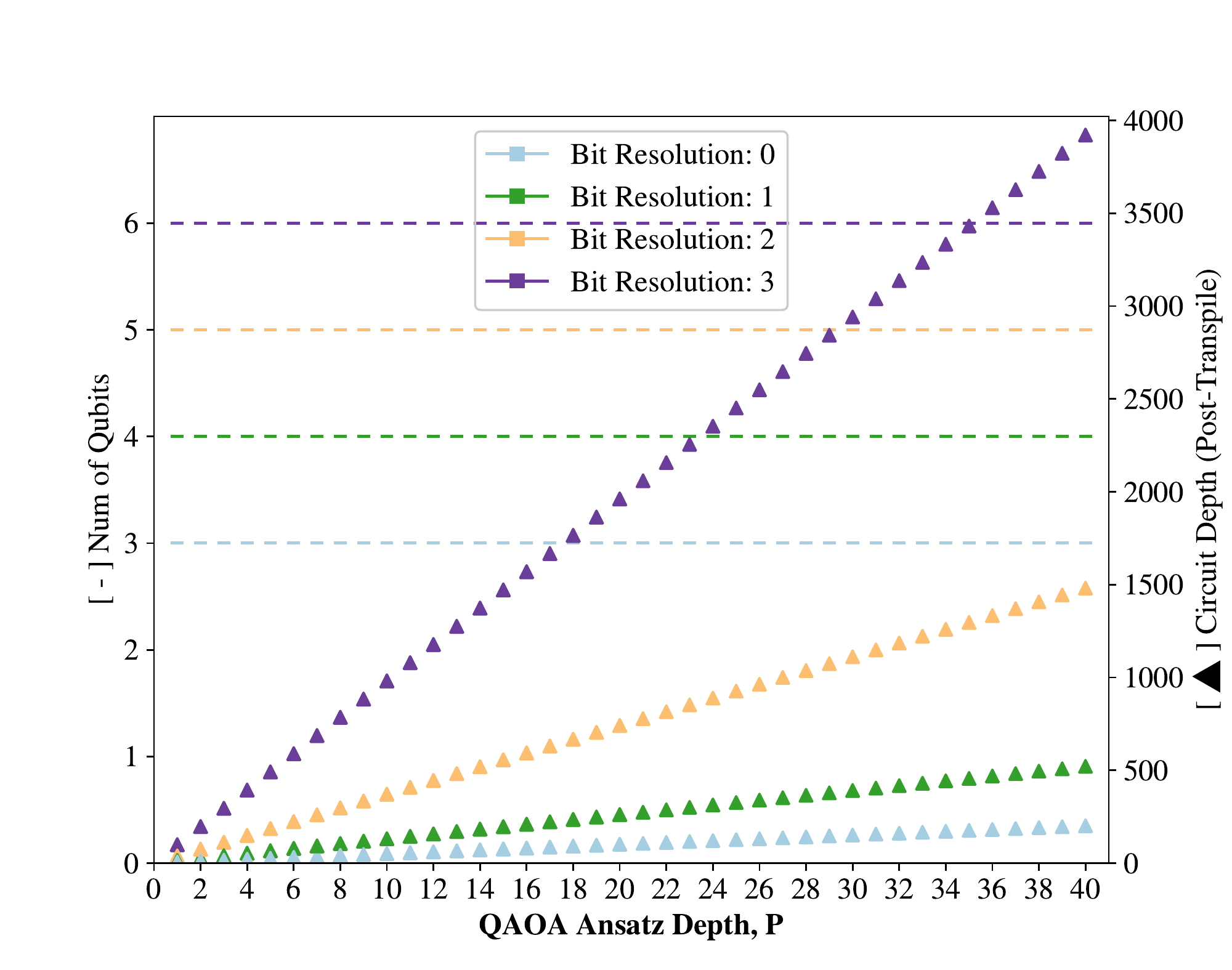}
        \caption{1D-ST, PUBO}
        \label{fig:QbitVDepthSTangPuboM1}
    \end{subfigure}
    \caption{Number of required qubits and circuit depth for different numbers of QAOA iterations $P$ and bit resolutions. The circuit depth is calculated after a transpilation targeted towards a gate set without multi-qubit gates beyond CNOTs.}
    \label{fig:Circuit-Width-and-Depth}
\end{figure*}

For the execution of the proposed approaches on real hardware, two criteria are essential: the circuit width (i.e., the number of qubits) and the circuit depth (i.e., the number of subsequent gate operations). Figure \ref{fig:Circuit-Width-and-Depth} exemplifies the both using the 1D-ST function, as it allows for a bigger scaling analysis in terms of bit resolution.

Before comparing the number of needed qubits for both approaches, we recall that the number of required qubits is entirely determined by the bit depth and the dimensions of the input domain, according to our chosen discretization. For PUBO, we can easily calculate the number of required qubits by adding up the number of bits used to represent each dimension of the input domain. For QUBO we can calculate this number by determining the number of required ancillary qubits and adding it to the number of qubits required for the PUBO formulation as a result of the quadratization. The number of ancillary qubits however relies heavily on the exact function and the techniques used for quadratization. We used a combination of different techniques based on the python package qubovert\footnote{\url{https://github.com/jtiosue/qubovert}} and boolean algebra simplifications. The resulting number of qubits for the 1D-ST function are displayed in figure \ref{fig:Circuit-Width-and-Depth}. Comparing the PUBO and QUBO approaches, we can clearly see a higher number of needed qubits in the QUBO variant, which gradually increases with bit resolution, as the number of qubic and quartic terms accumulate according to the chosen discretization, as described in section \ref{subsec:discretization}.

Continuing with the circuit depth (also displayed in figure \ref{fig:Circuit-Width-and-Depth}), we can observe a clear disadvantage of the PUBO approach when executed on a device that doesn't inherit a suitable gate set: While the QUBO's overall circuit depth at the highest complexities caps at less than 1400, the PUBO's overall depth reaches around 4000. The substantially higher circuit depth for current hardware raises an important potential drawback when deciding on whether to incorporate PUBO on current NISQ devices, where gate-fidelity is a significant constraint. For future quantum computers implementing suitable multi-qubit gates however, the scaling in terms of circuit depth would roughly equal that of the QUBO approach. Possibly, even less gates might be needed, as no interactions with ancillary qubits are needed. Another promising observation is the easier use of classical circuit simulators in PUBO, as they can generally provide arbitrary gate sets and thus allow for even shorter circuits while also needing fewer qubits.

\balance 

\section{Conclusion}
\label{sec:conclusion}
The conducted experiments clearly indicate that PUBO formulations achieve superior result quality over their quadratized QUBO analogues for continuous polynomial objective functions of a higher order. Until suitable multi-qubit gates become available, this manifests in a trade-off between the number of needed qubits (linearly higher for QUBO) and the circuit depth (linearly higher for PUBO). In terms of parameter training steps, both approaches performed equally. When using a quantum circuit simulator however, the wall-clock times for the PUBO formulations showed much better results, most probably because of the lower number of qubits that need to be simulated. For NISQ hardware, the performance difference is still mostly unclear and should be investigated in future work. We expect a strong dependence on the objective function, the input domain and bit resolution as well as their interplay with the error rates to be decisive. Finally, in the future we plan on exploring the combination of our findings with the existing positive results on using PUBO for combinatorial optimization problems to investigate the performance of PUBO formulations for NP-hard mixed integer problems.


\begin{acks}
This work was partially funded by the German BMWK project \textit{QCHALLenge} (01MQ22008A). The authors want to thank Johannes Kolb for his contributions to this research.
\end{acks}

\bibliographystyle{ACM-Reference-Format}
\bibliography{literature}




\end{document}